\documentclass[graybox]{svmult}

\pdfoutput=1
\usepackage{mathptmx}       
\usepackage{helvet}         
\usepackage{courier}        
\usepackage{type1cm}        
\usepackage{makeidx}         
\usepackage{graphicx}        
\usepackage{multicol}        
\usepackage[bottom]{footmisc}

\makeindex             

\usepackage{cite}
\usepackage{subfig}
\usepackage{stfloats}
\usepackage{url}

\begin{document}
%
\title*{The Ultrasound Visualization Pipeline - A Survey}

\author{\AA. Birkeland, V. \v{S}olt\'{e}szov\'{a}, D. H\"{o}nigmann, O. H. Gilja, S. Brekke, T. Ropinski and I. Viola}
\institute{\AA. Birkeland (Corresponding author) \at Department of Informations, University of Bergen, Norway, \email{asmund.birkeland@uib.no}}

\authorrunning{The Ultrasound Visualization Pipeline - A Survey, \AA. Birkeland et al.}%
\titlerunning{The Ultrasound Visualization Pipeline - A Survey, \AA. Birkeland et al.}%

\maketitle

\textbf{Note:} Also to appear in the Dagstuhl 2012 SciVis book by Springer. Please cite this
paper with its arXiv citation information.

\abstract*{
Ultrasound is one of the most frequently used imaging modality in medicine. The high spatial resolution, its interactive nature and non-invasiveness makes it the first choice in many examinations. Image interpretation is one of ultrasound's main challenges. Much training is required to obtain a confident skill level in ultrasound-based diagnostics. State-of-the-art graphics techniques is needed to provide meaningful visualizations of ultrasound in real-time. In this paper we present the process-pipeline for ultrasound visualization, including an overview of the tasks performed in the specific steps. To provide an insight into the trends of ultrasound visualization research, we have selected a set of significant publications and divided them into a technique-based taxonomy covering the topics pre-processing, segmentation, registration, rendering and augmented reality. For the different technique types we discuss the difference between ultrasound-based techniques and techniques for other modalities.}


\abstract{
Ultrasound is one of the most frequently used imaging modality in medicine. The high spatial resolution, its interactive nature and non-invasiveness makes it the first choice in many examinations. Image interpretation is one of ultrasound's main challenges. Much training is required to obtain a confident skill level in ultrasound-based diagnostics. State-of-the-art graphics techniques is needed to provide meaningful visualizations of ultrasound in real-time. In this paper we present the process-pipeline for ultrasound visualization, including an overview of the tasks performed in the specific steps. To provide an insight into the trends of ultrasound visualization research, we have selected a set of significant publications and divided them into a technique-based taxonomy covering the topics pre-processing, segmentation, registration, rendering and augmented reality. For the different technique types we discuss the difference between ultrasound-based techniques and techniques for other modalities.}


\section{Introduction}

Medical ultrasound has a strong impact on clinical decision making and its high significance in patient management is well established \cite{odegaard05, odegaard07}. Ultrasonography (US) has in comparison with CT, MRI, SPECT and PET scanning very favourable cost, great availability world-wide, high flexibility, and extraordinary patient friendliness. Despite these factors, ultrasonography stands out as the imaging method with the highest temporal resolution and also often the best spatial resolution. Furthermore, ultrasonography is a clinical method that easily can be applied bedside, even using mobile, hand-carried scanners \cite{gilja03} and even pocket sized scanners \cite{VScan}, thus expanding the field of applications considerably. However, low signal-to-noise ratio, "shadowing" and the relative small scan sector makes ultrasound images very difficult to interpret. Accordingly, improved visualization of the broad spectrum of ultrasound images has a great potential to further increase the impact of ultrasonography in medicine. 

As advancement of technology is fertilising and stimulating medical development, there is a continuous need for research and new applications in visualization. Visualization methods have the capacity to transform complex data into graphic representations that enhance the perception and meaning of the data \cite{gilja07}. Ordinary ultrasound scanning produces real-time 2D slices of data, and these dynamic sequences pose in itself a challenge to visualization methods. One example is functional ultrasonography (f-US), i.e. ultrasound imaging of (patho)physiology and/or organ function, in contrast to conventional imaging of anatomic structures. Using f-US, information on motility, biomechanics, flow, perfusion, organ filling and emptying can be obtained non-invasively 
\cite{gilja02, postema07}. 
Moreover, the 2D images can be aligned to form 3D data sets. 
In such cases, 3D visualization provides added value in terms of more holistic understanding of the data. Typical examples are demonstration of complex anatomy and pathology, pre-operative surgical planning or virtual training of medical students. Furthermore, there are now matrix 3D probes on the market that allow real-time 3D acquisition. To benefit from the high temporal resolution, advanced graphics techniques are required in ultrasound visualization, preventing the visualization technique from being the performance \emph{bottleneck}. This opens up new challenges to the visualization community to develop fast and efficient algorithms for rendering on-the-fly. 

In addition, co-registration techniques enable use of multi-modal data sets. Fusion imaging, where ultrasound is combined with either CT, MRI, or PET images, allows for more precise navigation in ultrasound-guided interventions. This challenging new arena demands advanced visualization research to enlighten how different data types can be combined and presented in novel ways.

The diversity of the ultrasound imaging technology provide a great tool for medical diagnostics, but the nature of the data can make it challenging to process. Techniques which work well for other modalities are being adapted to suit the special characteristic of ultrasound. In this paper we present an overview of the pipeline for advanced visualization specific to ultrasound data. The paper is divided into the chosen taxonomy, in essence each step of the visualization pipeline; pre-processing, segmentation, registration, rendering and augmented reality.

\section{Taxonomy}
 \begin{figure*}[b]
  \centering
  \subfloat[Heart]{\label{fig:region1}\includegraphics[height=0.9in]{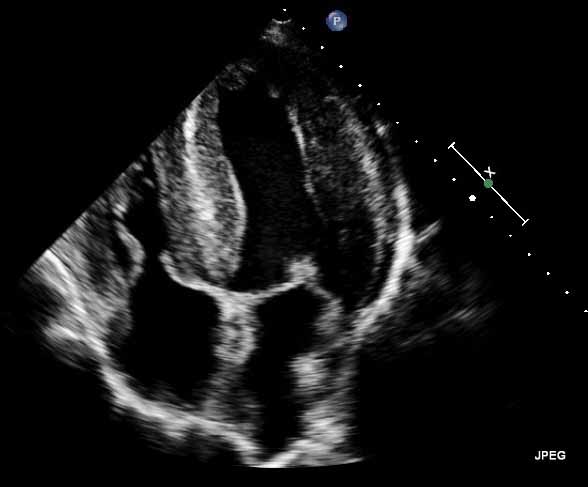}}
  \hspace{0.05in}
  \subfloat[Liver]{\label{fig:region2}\includegraphics[height=0.9in]{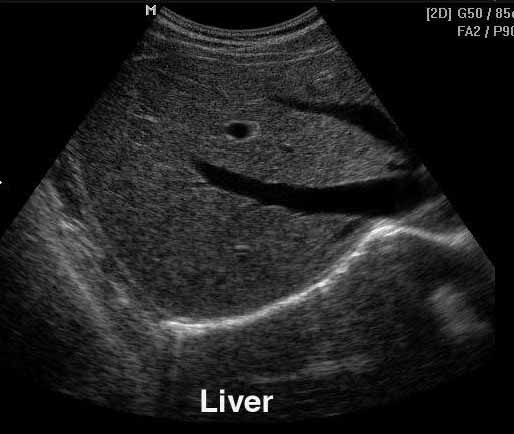}}
  \hspace{0.05in}
  \subfloat[Fetus]{\label{fig:region3}\includegraphics[height=0.9in]{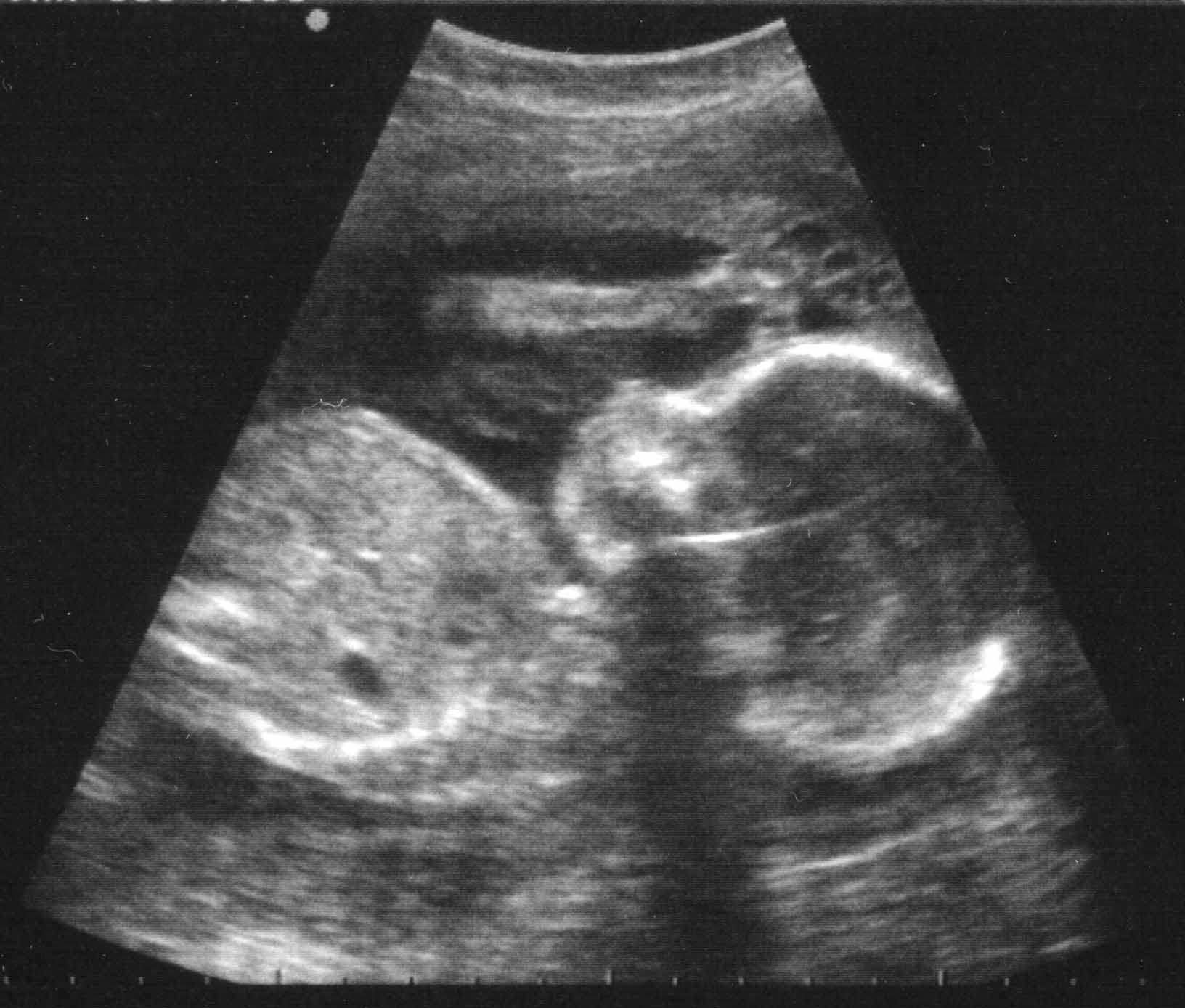}}
    \hspace{0.05in}
  \subfloat[Doppler]{\label{fig:doppler}\includegraphics[height=0.9in]{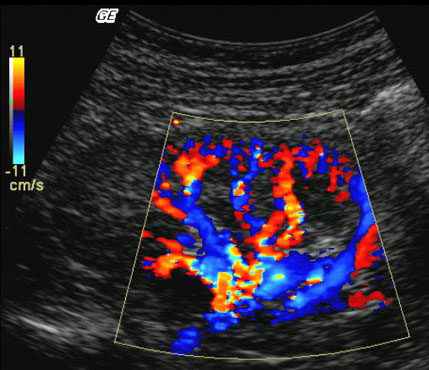}}
  \caption{Example ultrasound images from the cardiac~(a), gastric~(b), fetal~(c) and Blood flow~(d) domain.}
  \label{fig:anatomy}
\end{figure*}

Techniques for ultrasound visualization can be categorized in a variety of ways, e.g, when they where developed, what types of data modalities was utilized, which anatomy the technique was focused on, etc. The development of new ultrasound technology leads to different visualization techniques. The step from 2D ultrasound images to 3D freehand ultrasound (2D ultrasound with position information) revealed new challenges as spatial information could be included to generate volumetric data. The recent development of 2D matrix probes provided again a new challenge of 3D + time (4D) data visualization. Karadayi et al. published a survey regarding 3D ultrasound~\cite{karadavi09}. This paper has a greater focus on data acquisition and volume handling, but also give a brief overview over visualization of 3D ultrasound data.

\begin{figure*}[t]
	\includegraphics[width=\textwidth]{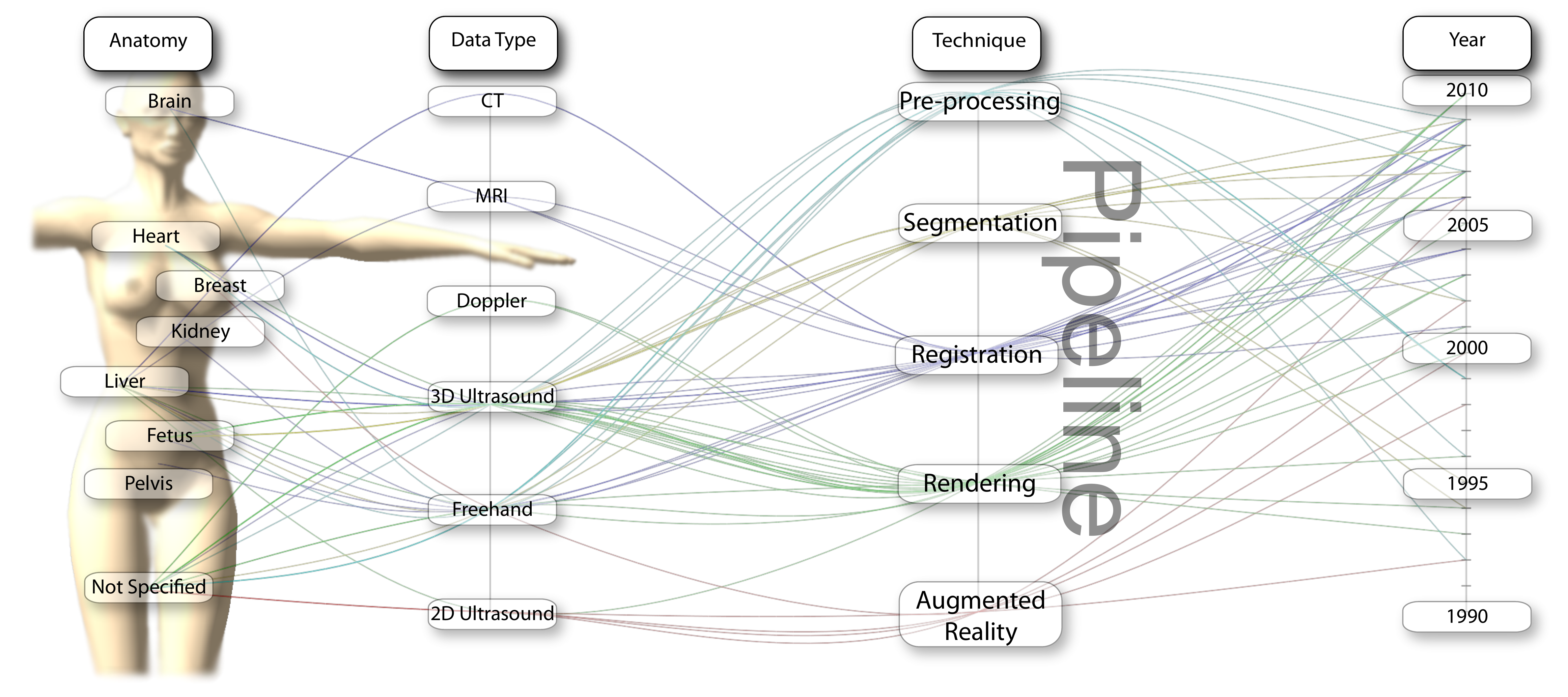}
	\caption{The different classifications shown in a parallel-coordinate plot. The colors depict which technique a publication has given the most weight.}
	\label{fig:usvisvis}
\end{figure*}
Another taxonomic scheme for ultrasound visualization is based on the different types of data the technique utilized. 3D Freehand and 4D ultrasound, pose very different challenges compared to 2D ultrasound or when handling multiple modalities. Blending b-mode ultrasound for tissue and color-doppler ultrasound for blood flow can be challenging enough in 2D if not in 3D. An example image is shown in Figure~\ref{fig:doppler}. In addition to the ultrasound input, the combination of other medical imaging modalities, such as CT or MRI with ultrasound, provide more information, but also more challenges to the visualization researcher. 

Different anatomic regions have different characteristics in ultrasound images, as can be seen in Figure~\ref{fig:anatomy}. For instance, in a liver scan one might look for tumors using a high-resolution abdominal 2D probe. For heart infarctions, the doctor might need to examine the strain in the heart muscle to detect defect muscle tissue. The wide spread in tissue and pathology difference lead to anatomically specific visualization techniques.

In this survey we categorized around 60 papers and from the different categories we generated a parallel-coordinate plot, show in Figure~\ref{fig:usvisvis}. Looking at the graph, we see an increase in rendering techniques for 3D ultrasound in the last five years. Volume rendering is often considered to be a \textit{solved} problem. However, our study shows that much work dealt with volumetric ultrasound data. Yet, 3D ultrasound rendering can still not be considered a solved problem. The high presence of noise, shadows from hyper-echoic areas and inconsistent data values provide a great challenge to make 3D ultrasound a more easy-to-use tool for examiners.

We also see an absence of augmented reality techniques for 3D ultrasound. Yet another trend is the neglecting of 2D ultrasound from the visualization community. 2D ultrasound is the most used modality by physicians and while presenting the signal data onto the screen is straight forward, understanding what you see is not so trivial. Increasing the readability of 2D ultrasound is mostly worked on in the commercial domain, aiming to give the company an edge over its rivals.

In Figure~\ref{fig:usvisvis} we see the categorized papers in a parallel coordinate plot where each axis corresponds to the different taxonomy classification. The second axis (the pipeline axis) is selected as the classification for this survey. Five categories where chosen based on what we recognize as the essential parts in the visualization pipeline for ultrasound data:

\begin{itemize}
	\item \emph{\textbf{Pre-processing}}:	Processing ultrasound data prior to segmentation, registration or rendering.
	\item \emph{\textbf{Segmentation}}:		Extracting features from ultrasound data.
	\item \emph{\textbf{Registration}}:		Combining ultrasound data with other types of medical imaging modalities.
	\item \emph{\textbf{Rendering}}:			Presenting ultrasound data.
	\item \emph{\textbf{Augmented Reality}}:	Combining ultrasound rendering with natural vision.
\end{itemize}

In the following sections we motivate need for each of the major topics and try to focus on significant techniques and how they deal with the characteristics of ultrasound data.

\section{Pre-processing}
3D ultrasound is often employed in clinical diagnostic imaging. If a dedicated 3D probe is unavailable, 3D volumes can be acquired using freehand ultrasound systems; a 2D probe with an attached tracking which places and orients the 2D images in 3D space. Volume compounding consists of two levels: acquisition and reconstruction. Precise reconstruction requires calibration of the tracking system and correction of pressure-induced artifacts from the probe onto the skin.

Ultrasound allows for extracting more information, such as tissue strain. Strain is a tissue-deformation property and can be used to detect functional deficiencies, e.g., from myocardial infarction. Strain determination via tissue tracking is a complex task and can be done by using tissue Doppler\cite{Heimdal98}. Deprez et al. advanced in 3D strain estimation by providing a better out-of-plane motion estimation~\cite{Deprez09}. Visualization of strain has however stagnated compared to the development of technology and is mostly depicted by elementary color coding.

For freehand ultrasound systems, it is necessary to calibrate the position and orientation of the 2D image with respect to the tracking sensor. Wein and Khamene proposed to make two perpendicular sweeps through tissue containing well-visible structures~\cite{wein08}. They used an optimization strategy to maximize the similarity between two volumes reconstructed from each sweep. 

To achieve the best possible quality of scans, the clinician press the probe against the body. However, the human factor causes a non-constant pressure and different deformations of underlying structures in the body. Prager et al. correlated images in the sequence and used a rigid translation in $x$ and $y$ directions followed by a non-rigid shift in depth $z$~\cite{prager02}.

Ultrasound acquisition takes place in polar coordinates $(\phi,R)$ for 2D or $(\phi,\psi,R)$ for 3D. The angles $\phi$ and $\psi$ correspond to the azimuth and elevation angles of the beam and $R$ is the depth of the tissue boundary which has reflected the echo. In order to use of-the-shelf 3D volume rendering techniques, the grid must be scan-converted to a Cartesian lattice. This can be done as a preprocessing step or on-the-fly directly at the rendering stage.

This section is dedicated to selected methods for \emph{volume reconstruction} from scan-converted freehand ultrasound and for \emph{data enhancement} tailored for ultrasound volumes, which in the pipeline typically follow the reconstruction stage. 

\subsection{Reconstruction}
\label{sec:prepros:rec}
Volume reconstruction from a set of 2D images needs to solve several important problems. Each images must be inserted precisely into the right spatial context, space-filling between individual images is also crucial and the high framerate of 2D ultrasound implies speed requirements.

A detailed categorization of reconstruction algorithms was done by Rohling et al.~\cite{rohling99-2} and Solberg et al.~\cite{Solberg07}. We adopt the categorization by Solberg et al.  into \emph{voxel-}, \emph{pixel-} and \emph{function-based} methods and complete it by recent works.

{\bf Voxel-based methods}, i.e., \emph{backward compounding}, run through the voxel grid and assign each of them a value estimated by an interpolation method such as the Stradx system~\cite{prager99}. It allows for real-time visualization of freehand ultrasound including plane re-slicing based on nearest-neighbour interpolation and later also for direct volume rendering~\cite{prager02}; They blend images generated by re-slicing as described in their previous work. Gee et al. also used nearest neighbor interpolation for direct plane re-slicing~\cite{gee04}. The reconstructed plane is intended for direct viewing - implying only one re-sampling stage. Linear, bilinear and trilinear interpolation methods have been also used~\cite{Thune96,Berg99}. Recent development by Wein et al. improve both quality and performance by applying a backward-warping paradigm implemented on dedicated graphics hardware~\cite{wein06}.

Karamalis et al. used interpolation on the GPU for high-quality volume reconstruction~\cite{karamalis2009}. They select an optimal orientation of reconstruction slices based on the orientation of the scans and reconstruct the volume by following this direction. Each sampling layer is reconstructed from scans which intersect this layer by interpolating intensity values between the intersections. The visualization pipeline includes two re-sampling steps: one during the reconstruction and one while volume rendering.

{\bf Pixel-based methods}, i.e., \emph{forward compounding}, traverse each pixel of all acquired 2D images and update the value of one or several voxels of the target grid. Gobbi and Peters used splatting as a high-quality interpolation method described a technique in real-time while the data is captured~\cite{gobbi02}.

{\bf Function-based methods} employ a specific function to interpolate between voxels. In most applications, the shape of the underlying data is not considered. Rohling et al. investigated the quality of interpolation using splines, which is a polynomial function~\cite{rohling99}. They compared this technique with other standard methods and showed that it produces more accurate reconstructions. 

{\bf Tetrahedron-based methods} reconstruct 3D model built from tetrahedra using an iterative subdivision of an initial tetrahedron instead of a regular grid~\cite{roxborough00}. The subdivision terminates, if all tetrahedra contain one data point. Each point is assigned a value which corresponds to the barycentric coordinates of the data point in this tetrahedron. This strategy is adaptive; the model adapts as new data is streamed in.

We listed selected algorithms in categories based on how they were implemented. If choosing a specific algorithm, one must choose between speed and quality. Solberg et al. compared the performance of some of the algorithms~\cite{Solberg07}. From all listed methods, the radial-based function reconstruction by Rohling et al.~\cite{rohling99} delivers reconstructions of the best quality but it is also the most computationally costly. However, the increasingly powerful dedicated graphics hardware for computational acceleration solves this problem.

\subsection{Data Enhancement}

Ultrasound is a challenging modality for visualization due to its natural properties such as low dynamic range, noisiness and speckle~\cite{sakas95}. Also, the geometric resolution varies with depth and the tissue boundaries can be several pixels wide depending on their orientation. Tissue boundaries can even disappear if they are parallel to the ultrasound beam. 2D images are preferred without filtering and enhancement. Speckle patterns refer to the texture of the tissue boundary which is a valuable information for clinicians. However, speckle in 3D brings no added value to the visualization and is considered as an artifact same as noise. Therefore, prior to the rendering stage, the 3D data is filtered to enhance its quality. We present speckle reduction techniques separately and followed by dedicated preprocessing for specific applications. 

{\bf Speckle reduction.} For a review on early speckle reduction techniques, refer to the survey of Forsberg et al.~\cite{forsberg91}. Belohlavek et al.~\cite{Belohlavek92} uses the \emph{eight hull} algorithm with a geometric filter~\cite{Crimmins85}. Recent techniques are based on region growing~\cite{chen03}, adaptive filtering~\cite{Shankar06}, compression techniques~\cite{gupta05} and anisotropic diffusion filters~\cite{krissian07}.

{\bf Application-dedicated enhancement.} Systems usually employ a blend of image-processing techniques to enhance the data. Sakas et al. listed techniques with a good trade-off between loss of information and quality~\cite{sakas95}. They employed Gaussian filters for noise reduction, speckle-removal methods for contour smoothing and median filters for gap closing and noise reduction. Median filters remove small surface artifacts and preserve the sharpness of boundaries. There exist fast implementations where a histogram can be used to keep track of values~\cite{huang79}. Still, they require a more advanced memory management, making them less parallelizable than the evaluation of fast Gaussian filters. Lizzi and Feleppa described a technique to increase the axial resolution by processing the signal in the frequency domain. This resolution gain is especially valuable in opthalmology when visualizing thin layers within the cornea~\cite{Lizzi00}.

\section{Segmentation}
Selecting interesting features to be visualized is important to be able to root out the occluding elements from large datasets.

For most modalities, segmentation can be performed by extracting regions with similar data values. For instance, because of the physical properties of x-rays, the data values in a CT scan are recorded into Hounsfield units which provide a good basis for binary thresholding techniques for certain tissue types. 
Early work indicated that binary thresholding techniques are not very well suited for ultrasound data~\cite{steen94}. More sophisticated techniques are required for satisfactory segmentation. An extensive survey on ultrasound image segmentation was been presented by Noble and Boukerroui \cite{noble06} in 2006. In this section we have focused on significant publication from recent years.

\begin{figure}[t]
	\includegraphics[width=0.9\linewidth]{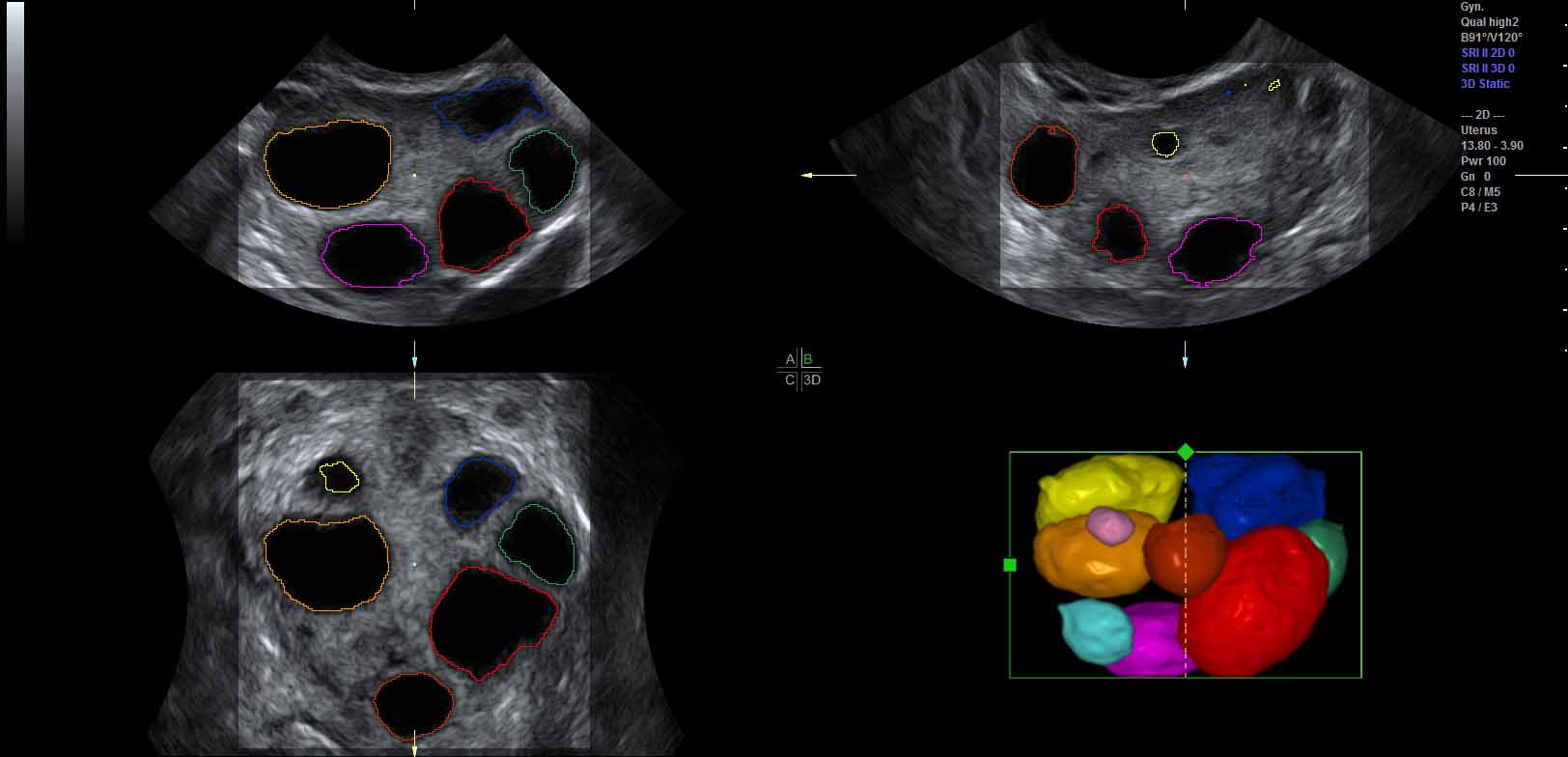}
	\caption{Automatic segmentation of the ovarian follicles \cite{deutch09}.}
	\label{fig:segmentation}
\end{figure}

To increase robustness of the ultrasound segmentation, the various approaches are usually tailored for specific anatomies. Carneiro et al. have developed an automatic technique for segmenting the brain of a fetus~\cite{carneiro08}. By first detecting the cerebellum, the system can \emph{narrow down} the search for other features. On the other hand, segmentation is an extremely critical procedure which may obscure diagnostically relevant aspects of the anatomy under examination. Consequently, fully automatic segmentation techniques have not been implemented in clinical systems so far, with the exception of a method for follicle volumetry \cite{deutch09}, as shown in figure~\ref{fig:segmentation}.
A great challenge with ultrasound segmentation is that the data is dependent on many factors. For one, different positions and orientations of the probe, while looking at the same anatomical part, can provide very different images. Hyper-echoic regions cast shadows onto the tissue behind it according to the probe position. This alone, makes ultrasound segmentation data highly uncertain. Most segmentation techniques return a model with no indication of the uncertainty of the result. To compensate for the fuzzy nature of the ultrasound data, Petersch et al. developed a soft segmentation technique for 3D ultrasound data~\cite{petersch06}. This technique calculate a probability map for 3D ultrasound data, which in turn can be used to create \emph{soft} representations of the features extracted.


\subsection{Clipping}
\begin{figure}[t]
	\centering
	\includegraphics[width=\linewidth]{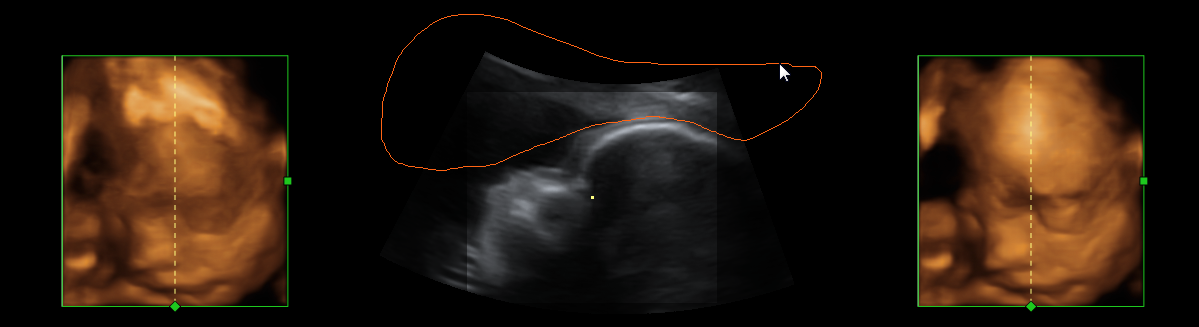}
	\caption{Using MagiCut to Clip the volume generating a clear view to the desired structure \cite{magicut99}.}
	\label{fig:clipping}
\end{figure}

Feature extraction can be computationally costly. In-vivo 3D ultrasound examination cannot always afford the extra time necessary to extract the interesting structures. Therefore clipping is commonly used tool in live visualization of 3D ultrasound. Interactively removing regions which are not interesting, the user gets a clear view of the features normally occluded. 
Sakas et al. developed a clipping tool in their ultrasound visualization system~\cite{sakas95} which is nowadays a standard feature in commercial 3D ultrasound systems. The user can in-vivo segment the dataset using three approaches. Drawing on one of the three axially-aligned slices, selecting everything along the current axis and within the sketch. Another tool is based on sketching directly on the 3D rendered scene. Each voxel is the projected onto the screen and removed if it lies within the marked area. The third clipping tool is based on the distance from a single mouse-click on the view-plane. A hemispherical wave front is propagated from the seed-point and stops when the voxels reach user-specified threshold. Figure~\ref{fig:clipping} show an example of clipping implemented in the GE Voluson machines \cite{magicut99}.

\section{Registration}
Merging ultrasound with other modalities can be very beneficial. While ultrasound provides high resolution images at a high frame-rate, other modalities, such as MRI or CT can provide information complimentary to the ultrasound images. Data registration is the process of transforming different modalities into the same reference frame to achieve as much comprehensive information about the underlying structure as possible. While CT and MRI are typically pre-operative imaging techniques, ultrasound can easily be performed live during surgery. For instance, the radiation from CT is dangerous and the large electro magnets in a MRI scanner require that everything in the room is non-magnetic. Recently Curiel et al. built a non-magnetic ultrasound scanner for proper simultaneous intra-operative imaging~\cite{curiel07}. There where some electric interference between the two modalities. Yet, the technique is promising, although availability will most likely be very low.

\begin{figure}[tb!]
  \centering
  \subfloat[]{\label{fig:registration}\includegraphics[height=1.2in]{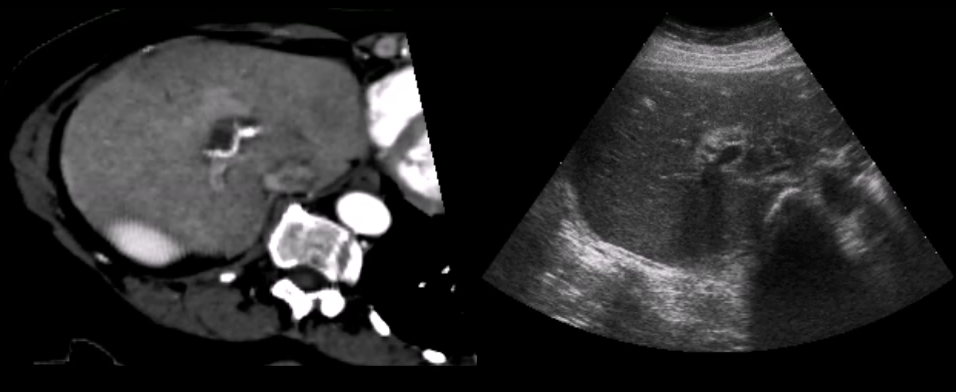}}
  \hfill
  \subfloat[]{\label{fig:cutaway}\includegraphics[height=1.2in]{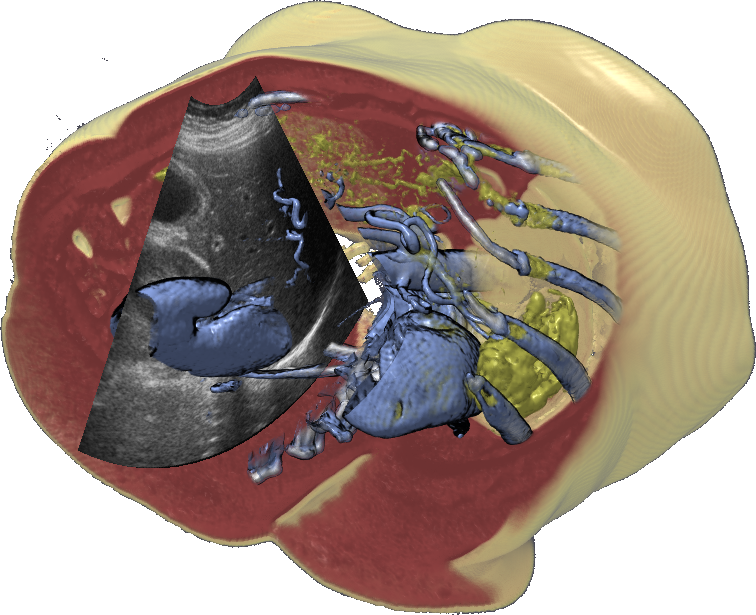}}
  \caption{Registering ultrasound and MRI, se Figure~(a), enables advanced visualization techniques to provide a better structural overview~\cite{burns07}, as shown in Figure~(b).}
  \label{fig:reg}
\end{figure}

Nikas et al. published an evaluation of the application of co-registered 2D ultrasound and MRI for intra-operative navigation~\cite{nikas03}. Ultrasound based navigation shows promising results due to live acquisition at high frame rates and easy portability.
For prostate brachytherapy a combination of ultrasound and co-registered CT can be used, as shown by Fuller et al.~\cite{fuller05}. Existing commercial products apply optical tracking for intra-operative navigation during neurosurgery \cite{sonowand11}.
Figure~\ref{fig:reg} shows how ultrasound and MRI can be blended together into a single reference frame \cite{burns07}. 



Registration can be divided into two different types: Rigid and non-rigid. 
Rigid registration can be used to quickly obtain a registration between two modalities and is suitable for rigid anatomies such as the skull. 
A common approach to register two images is to search for the transformation which minimize a difference function, for instance sum-of-square-difference. Direct image based registration between ultrasound and CT or MRI can be difficult due to the different nature of the imaging techniques and usually some pre-processing, such as filtering, is required. For instance, an approach presented by Leroy et al. used a gradient-preserving speckle filter and then looked for the similarity in the gradients.

Penney et al. proposed a technique for registering MRI and ultrasound. The system calculates a probability map of each element being a part of a liver-vessel~\cite{penney04}.
Later Penney et al. extended their technique for CT-ultrasound registration of the pelvis and femur~\cite{penney06}. The system was validated using cadavers, showing that the registration was accurate to a 1.6mm root-mean-square error on average.
A similar technique for the cardiovascular domain was proposed later by Zhang et al.\cite{zhang06}.

Combining segmentation with registration, King et al. presented a technique for registering pre-segmented models with ultrasound~\cite{king09}. The technique predicts the probability that the ultrasound image was produced by the segmented anatomy. 

In addition to a rigid transformation, affine registration includes non-uniform scaling which sometimes needs to be applied in order to get a more correct registration. 
Wein et al. developed an automatic affine-registration technique between CT and ultrasound\cite{wein08}. To provide a better similarity of the ultrasound an CT, the system creates a simulated ultrasound image out of the CT scan based on the tracked probe position. The simulated ultrasound image is generated using a ray-traced approach to calculate the ultrasound wave reflection and attenuation in the tissue. To simulate tissue specific echogeneity, they apply a angle-independent polynomial function based on which tissue the region corresponds to.

External pressure or different laying positions of the patient when acquiring the images. To account for local deformations while imaging soft tissue, a more complex registration is required. 
Papenberg et al. proposed two approaches for CT ultrasound registration~\cite{papenberg08} given a set of paired landmarks in both the CT and ultrasound data-set. One approach use the landmarks as hard constraints and in the other, the landmarks are considered as soft constraints and are combined with intensity value information, in this case the normalized gradient field. The paper shows a non-rigid registration between the liver vascular structures. The latter technique was later evaluated by Lange et al.~\cite{lange08}.

\section{Rendering}
Visual presentation of the data is the last stage of the pipeline before the user. The basic B--mode ultrasound images can be depicted on a screen in a straight-forward manner as varying pixel intensities according to the echo amplitude. Doppler information can be included as well with color-encoded blood-flow direction. Other data, such as tissue strain can also be included into 2D as overlays. Another example of overlays is the \emph{CycleStack Plot} which superimposes the respiratory signal onto a selected feature of interest in the ultrasound image~\cite{lee10}. Doctors use this information to account for the respiration-caused motion of the tumor in order to minimize the damage done by certain tumor treatments.


\textbf{Freehand ultrasound} In Section~\ref{sec:prepros:rec}, we discussed how freehand ultrasound systems can be used to create large volumes by putting images into 3D spatial context.  Garrett et al. presented a technique for correct visibility ordering of images using a binary positioning tree~\cite{garrett96}. Visualization of large volumes leads to visual clutter. Therefore, Gee et al. extended existing re-slicing tools to create narrow-band volumes which contain less elements and are easier to present~\cite{gee02}. 

\textbf{3D ultrasound} is not as trivial to present due to its natural properties. In an early work, Nelson and Elvis discussed the effect of existing techniques for presenting 3D ultrasound data, such as surface fitting and volume rendering~\cite{Nelson93}. Later, seven ultrasound-dedicated volume projection techniques were evaluated by Steen and Olstad \cite{steen94}. They included maximum intensity projection (MIP), average intensity projection (AIP) and gradient magnitude projection (GMP). The techniques were applied to 3D fetal data, where GMP was valued to give the best detail and robustness towards viewing parameters.

Data definition in polar coordinate system is another challenge for ultrasound volume rendering. Kuo et al. presented a technique for quick on-the-fly scan-conversion~\cite{kuo07}. The reduce the costs of the functional evaluation of $tan(\phi)$ and $tan(\psi)$, the functional values were pre-calculated and stored in a texture as a look-up-table. 

\textbf{Surface Rendering} is a common tool for many imaging modalities. In ultrasound, the low signal-to-noise ratio and parallel tissue boundary discontinuities make defining smooth surfaces difficult. Smoothing of a surface can be performed at the rendering stage. Fattal et al. presented an approach to render smooth surfaces from 3D ultrasound~\cite{fattal01}. The surface is extracted based on the variational principle. Fuzzy surface rendering is done by a technique called oriented splatting. Oriented splatting creates triangles aligned with the gradient of the surface function, the triangle is then colored with a Gaussian function and rendered in a back-to-front order. 
Wang et al. proposed an improved surface rendering technique for 3D ultrasound data of fetuses \cite{wang08}. To remove the noise and to preserve edges, a modified anisotropic diffusion is first applied to the dataset. To enhance low intensities which appear due to signal loss as the sound wave propagates through the tissue, a light absorption function based on the distance from a point is applied to the data. Finally, a texture-based surface rendering is used, where the texture is extracted from images of infants. The textures are warped and blended with the surface of the fetus face. 
To create smooth surfaces and remove unimportant noise in direct volume rendering, Lim et al. proposed a filtering technique in their GPU based ultrasound rendering framework \cite{lim09}. This technique employs different sized filters to smooth out the noise. 

\subsection{Transfer Function Design}

For direct volume rendering, {\em transfer functions} map ultrasound data, i.e., voxel echogenicity in B--mode imaging, and frequency information in Doppler imaging, onto colors and opacities. Usually, this mapping is based on look-up tables. In color Doppler imaging the commonly used red-to-blue color transfer function encodes direction and velocity of flow, whereas a variety of predefined color maps is in use for B--mode volume rendering. Custom color map editors are available, but hardly ever used. Overall, there is a well-established set of color-maps used in clinical practice.

Different from color transfer functions, where the selection largely depends on the preferences of the sonographer, the proper design of an appropriate {\em opacity transfer function} (OTF) is crucial: When designing OTFs, the goal is to assign a high opacity to voxels of structures of interest, while mapping all other samples to low opacities, thus avoiding any occlusion of the target structure. Whereas computed tomography allows classification of tissue based on voxel intensities, tissue classification-based transfer functions do not work in B--mode imaging due to the completely different data characteristics: Generally, a high signal intensity arises at a transition {\em between} tissues of different acoustic properties. Thus, at least in the case of soft tissue structures, we will measure high signal intensity at transitional areas and lower intensity signals within homogeneous tissue. This is the reason for applying monotonically increasing opacity transfer functions in DVR of ultrasound data: The aim is to opacify the tissue transitions in the hope of obtaining a visualization of an entire target structure. 

\begin{figure}[tb!]
  \centering
  \includegraphics[width=0.8\linewidth]{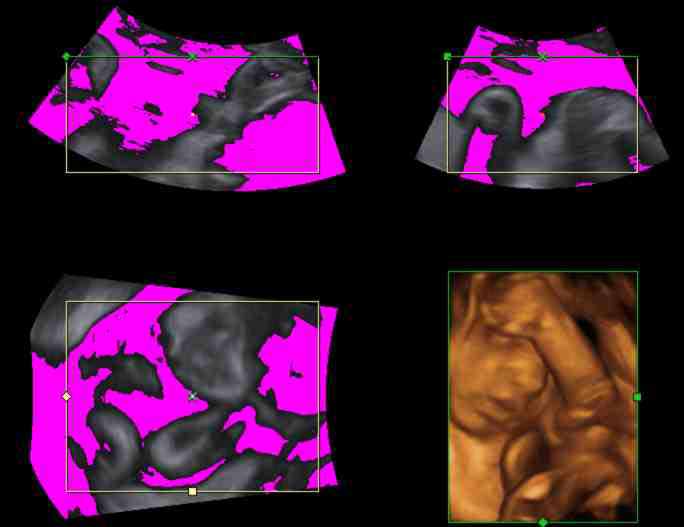}
  \caption{The parameter $I_{thresh}$ determines which echo intensity values to render transparent. A user control with immediate feedback, indicating transparent regions in pink, is essential.}
  \label{fig:otf_threshold}
\end{figure}

The most commonly used OTF in volume rendering of B--mode data assigns voxels to one of three classes depending on their echogenicity, namely invisible, transparent, and opaque. 
The corresponding piecewise linear OTF is modified manually by means of two parameters, namely a threshold intensity $I_{tresh}$ and a transparency value $\alpha$ controlling the increase of opacity for intensities above $I_{thresh}$. The effect of modifying $I_{thresh}$ is depicted visually on MPR images, see Fig.\ref{fig:otf_threshold}.

The parameters of the OTF affect the rendered image in a substantial way: The lower $I_{thresh}$, the lower the rendered image's brightness, due to an increasing number of hypoechoic voxels contributing to the image. Furthermore, the OTF affects depth contrast, i.e., the contrast arising from a spatial discontinuity in the target structure, and tissue contrast, i.e., contrast due to different echogenicity of adjacent tissue. See \cite{honigmann03} for an evaluation of these effects on linear and parabolic OTFs. On the other hand, any modification of fundamental acquisition parameters, such as, e.g., overall gain, or depth gain compensation, and any change of the position of the transducer or the target structure, changes the echogenicity distribution and thus requires modifying the OTF for an optimum visualization. Obviously, for a real time imaging modality incessant modification is not feasible. Hence, in clinical practice sonographers use a default OTF providing reasonable visualization in the majority of cases, and hardly ever touch the OTF control panel. 

Therefore, there is a need for automatic determination of an optimal OTF for every single acquisition. Due to the distinct characteristics and the real-time nature of ultrasound imaging, most conventional approaches for transfer function design have proven inadequate or require substantial modification in order to be applicable to ultrasound volume imaging. Among the most important advances in transfer function design for CT data is the work by Kindlmann et al. \cite{KINDLMANN1998} and subsequent work by Kniss et al. \cite{KNISS2002}, introducing the concept of histogram volumes for semi-automatic generation of OTFs for datasets where the regions of interest are boundaries between materials of relatively constant value. In \cite{JAN2009}, von Jan et al. adapt this approach to ultrasound data and apply it successfully to 3D freehand acquired volumes of  hyperechoic structures. 

H\"{o}nigmann et al. suggest an approach dedicated to the specific problem of rendering hyperechoic structures embedded in hypoechoic fluid \cite{honigmann03}. By analyzing so called {\em tube cores} they yield an estimate for the position of the most prominent tissue transition, in rendering direction. Voxel intensities {\em prior to} and {\em at} the detected interface steer the extent of modification of an initial, parabolic OTF in a multiplicative way. 
In a subsequent publication the authors assess the temporal coherence of the tube core method and conclude that it is sufficiently efficient and robust for online-computation of OTFs for an entire sequence of acquired volumes, if smoothing in the temporal domain is employed\cite{petersch05}. 

Additional challenges arise when it comes to DVR of multiple datasets.

\subsection{Multi-Modal Rendering}
Multi-modal rendering is meant to bring two or more data-sets of the same object, into a single image. Having two or more datasets in the same scene creates a challenge to keep the cluttering of less interesting regions to a minimum from the datasets. For ultrasound, 3D Doppler data can be acquired simultaneously with 3D B--mode data. 
Jones et al. discuss several approaches to explore and visualize 4-D Doppler data~\cite{jones03}. Multi-planar rendering, showing several slices at once, surface fitting of the Doppler data based on YCbCr color scheme values to improve separation between Doppler data and B--mode data.  An approach is presented to blend multi-planar slice rendering into a DVR scene. The DVR is shown very transparent and the slices provide better detail along the perspective.
A different way of combining B--mode with Doppler data was presented by Petersch and H\"{o}nigmann~\cite{petersch07}. They propose a one level composite rendering approach allowing for blending flow and tissue information arbitrarily. Using silhouette rendering for the B--Mode and a mix of Phong shaded DVR and silhouette rendering on color Doppler.  

A new technique for blending Doppler and B--mode was introduce by Yoo et al.~\cite{yoo07}. Instead of blending two 2D rendered images (post fusion), or a blending the two volumes while rendering (composite fusion), it proposes a way to do both called progressive fusion (PGF). Post fusion has a problem with depth blending and composite fusion will get a too early ray termination. PGF compensate for this by using an if-clause to adjust the alpha-out value in the ray-caster to either the Doppler-signal or the B--mode-signal.

Burns et al. applied illustrative cut-aways combined with 3D freehand ultrasound \cite{burns07}. This provide a better spatial overview for the ultrasound images.
To add more information onto the 2D ultrasound image, Viola et al. proposed an approach to enhance the ultrasound image by overlaying higher order semantics\cite{viola08}, in this case in the form of Couinaud segmentation. The segmentation is pre-defined in a CT dataset and visually verified using exploded views. To combine it with ultrasound images, CT dataset is co-registered with the ultrasound using rigid transformation according to user defined landmarks. The different segments are superimposed onto the ultrasound image enabling the user to directly see which segments are in the visible. To improve ultrasound video analysis, Angelelli et al. used a degree-of-interest (DOI) function superimposed on the image \cite{angelelli10}. The video sequence was presented as a graph, where the height was defined by the amount the current ultrasound image covered the DOI-function.

\subsection{Shading and Illumination}
Light is an indispensable part of scenes we see in real life. Also in computer graphics, light sources and light transport models have to be taken into account, when rendering realistic scenes. In volume graphics, the problem of illumination and light transport has been tackled by a handful of researchers as well. 

We distinguish between local and global illumination models. Local illumination models use gradients of the volumetric function instead of surface normals to evaluate the diffuse and specular terms of the Phong illumination model~\cite{levoy88isosurface}. While local illumination models already reveal structures, global illumination methods result in a more real appearance, which further supports spatial perception. While gradient-based local illumination methods are faster to evaluate, gradient computation is sensitive to noise and high frequencies, which are natural properties of ultrasound data. 

\begin{figure}[tb!]
  \centering
  \subfloat[]{\label{fig:chromadepth}\includegraphics[height=1.3in]{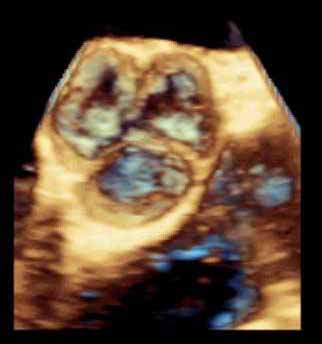}}
  \hspace{0.05in}
  \subfloat[]{\label{fig:hearttimo}\includegraphics[height=1.3in]{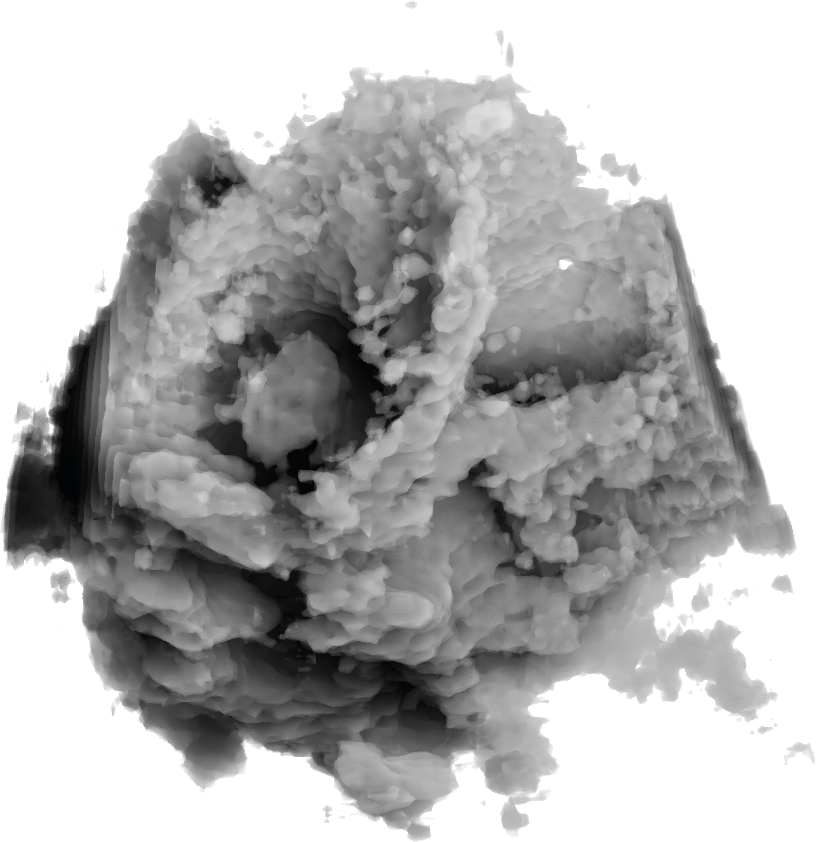}}
  \hspace{0.05in}
  \subfloat[]{\label{fig:heartme}\includegraphics[height=1.3in]{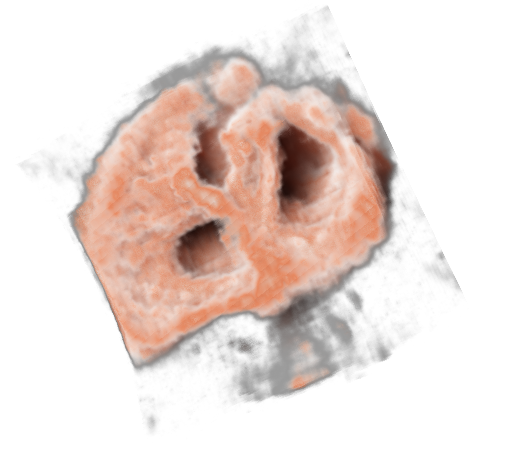}}
   \caption{(a): Diastole of the aortic valve on a modern ultrasound workstation using color-coding based on depth. (b): Rendering of 3D ultrasound of human heart with shadowing from the work of Ropinski et al.~\cite{Ropinski10} and~(c) rendered using the technique presented in the work of \v{S}olt\'{e}szov\'{a} et al.~\cite{solteszova10}.}
  \label{fig:hearts}
\end{figure}

\begin{figure}[tb!]
  \centering
  \subfloat[Phong]{\label{fig:comparison:phong}\includegraphics[width=0.28\linewidth]{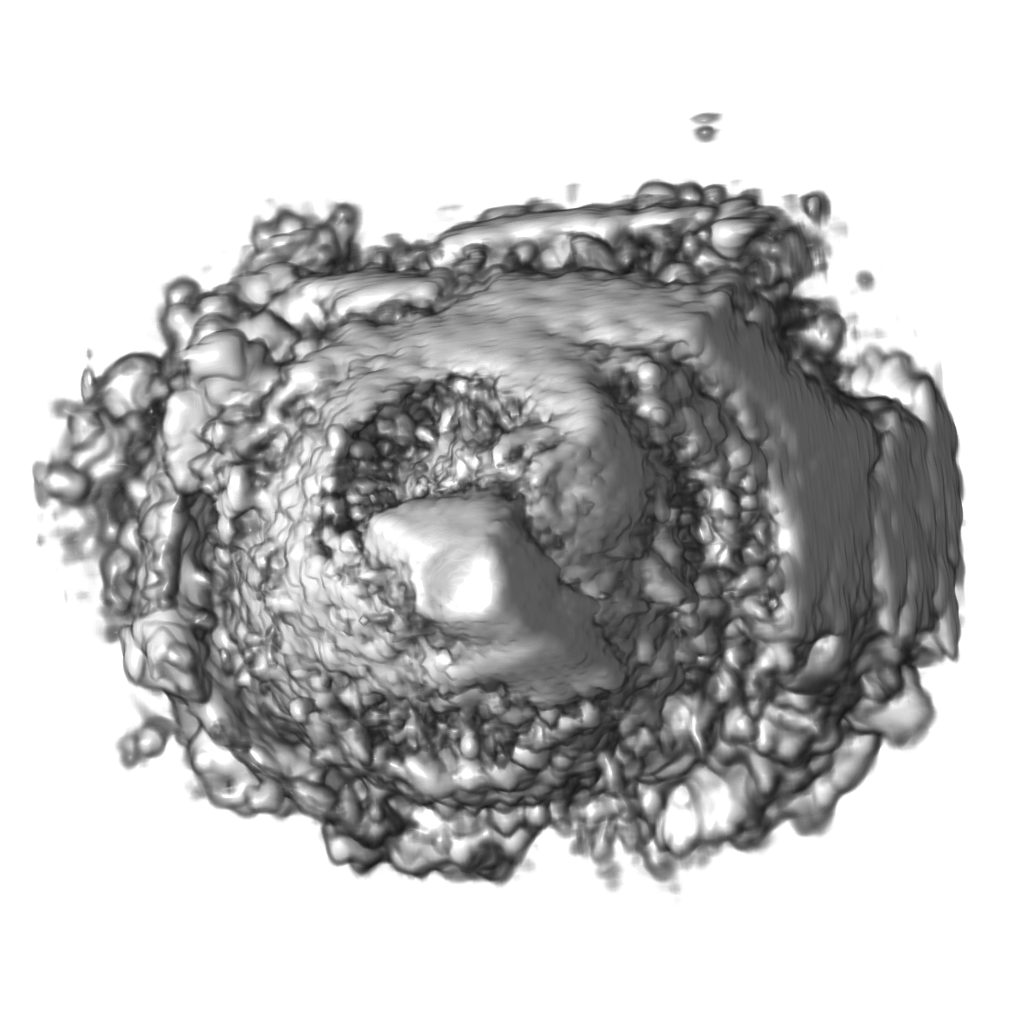}}
  \subfloat[\cite{Ropinski10}]{\label{fig:comparison:shadow}\includegraphics[width=0.28\linewidth]{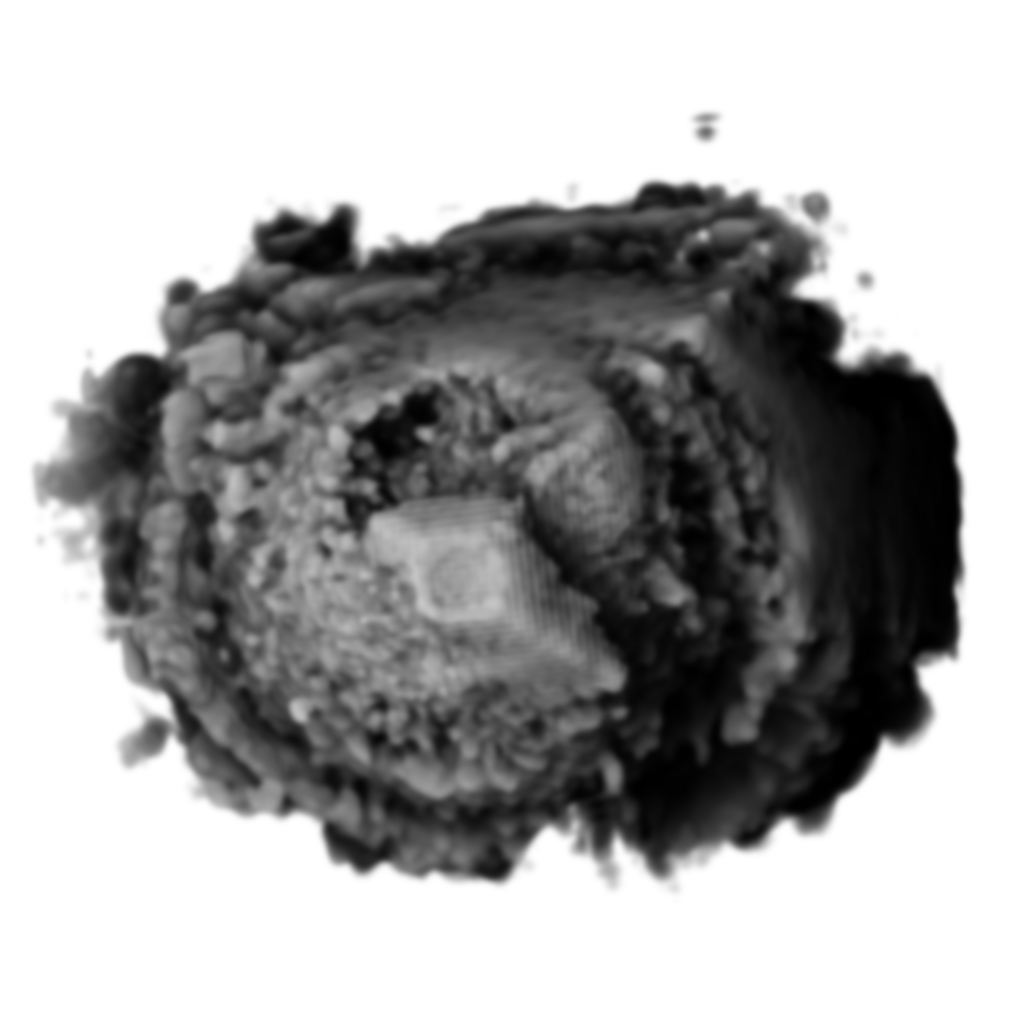}}
  \subfloat[\cite{solteszova10}]{\label{fig:comparison:multidir}\includegraphics[width=0.28\linewidth]{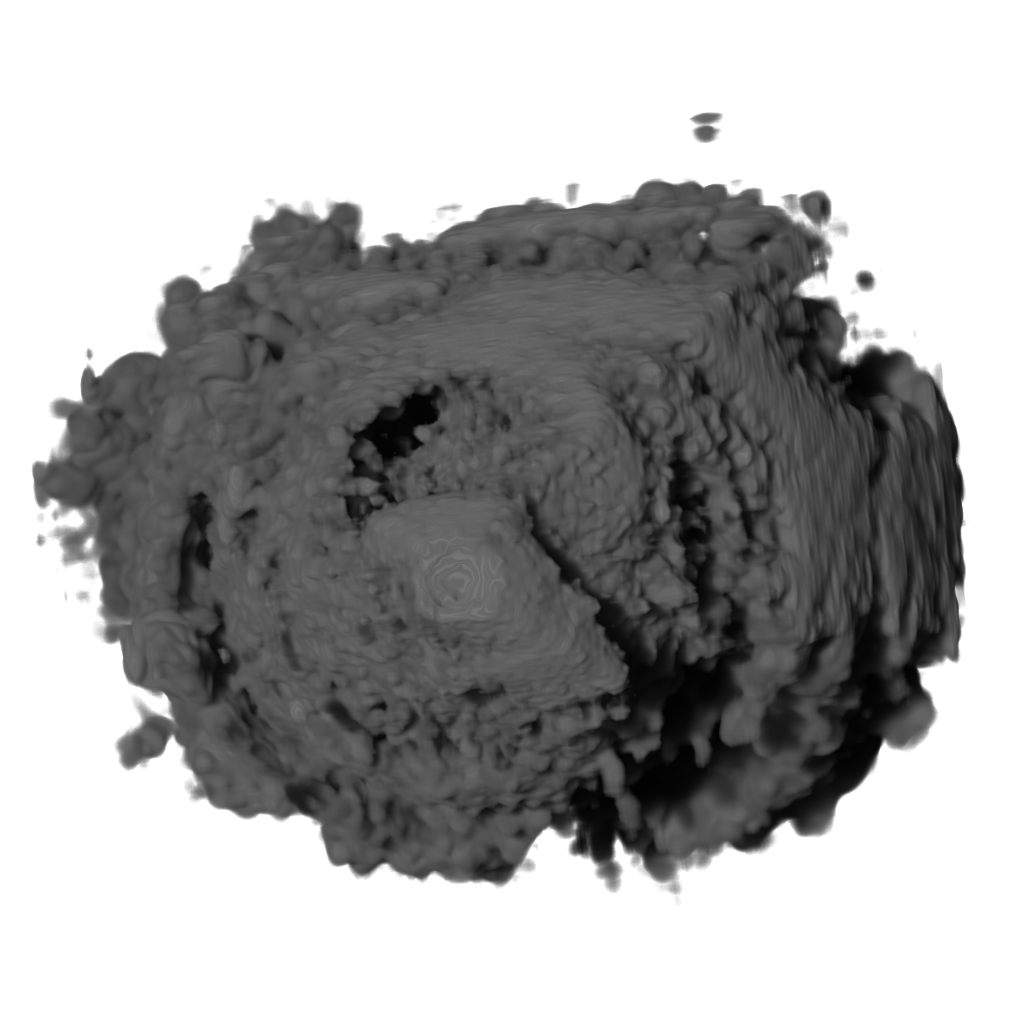}}\\
  \subfloat[\cite{kniss02halfangle}]{\label{fig:comparison:half}\includegraphics[width=0.28\linewidth]{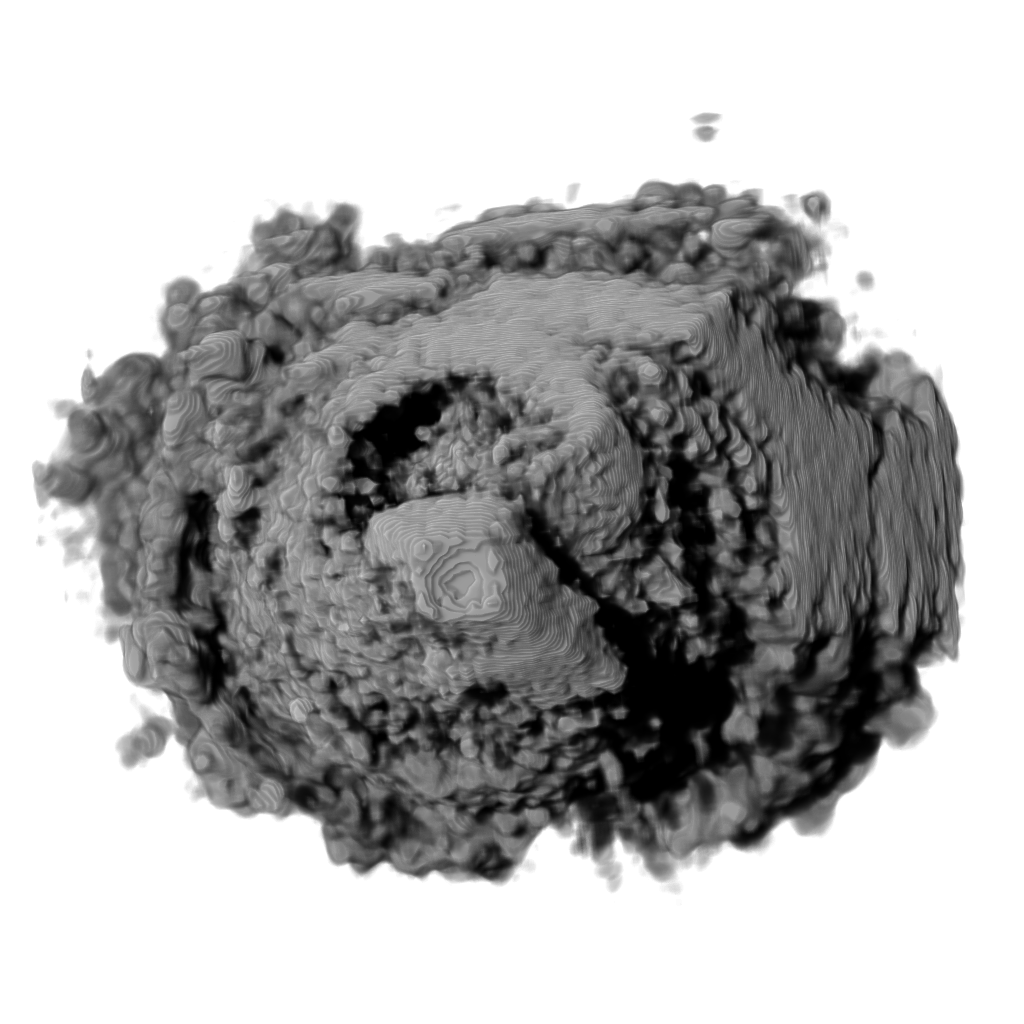}}
  \subfloat[\cite{schott09directionalocclusion}]{\label{fig:comparison:dirocc}\includegraphics[width=0.28\linewidth]{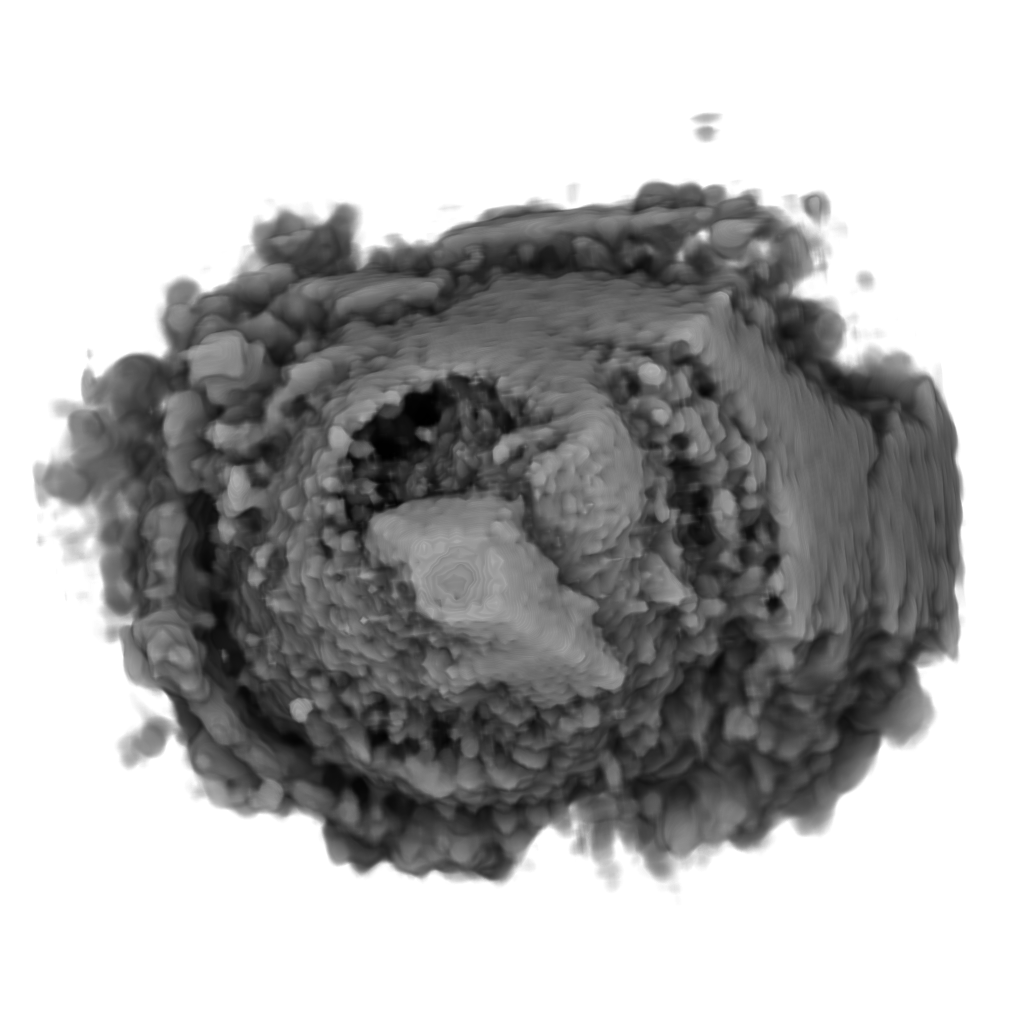}}
  \subfloat[\cite{lindemann10materials}]{\label{fig:comparison:sh}\includegraphics[width=0.28\linewidth]{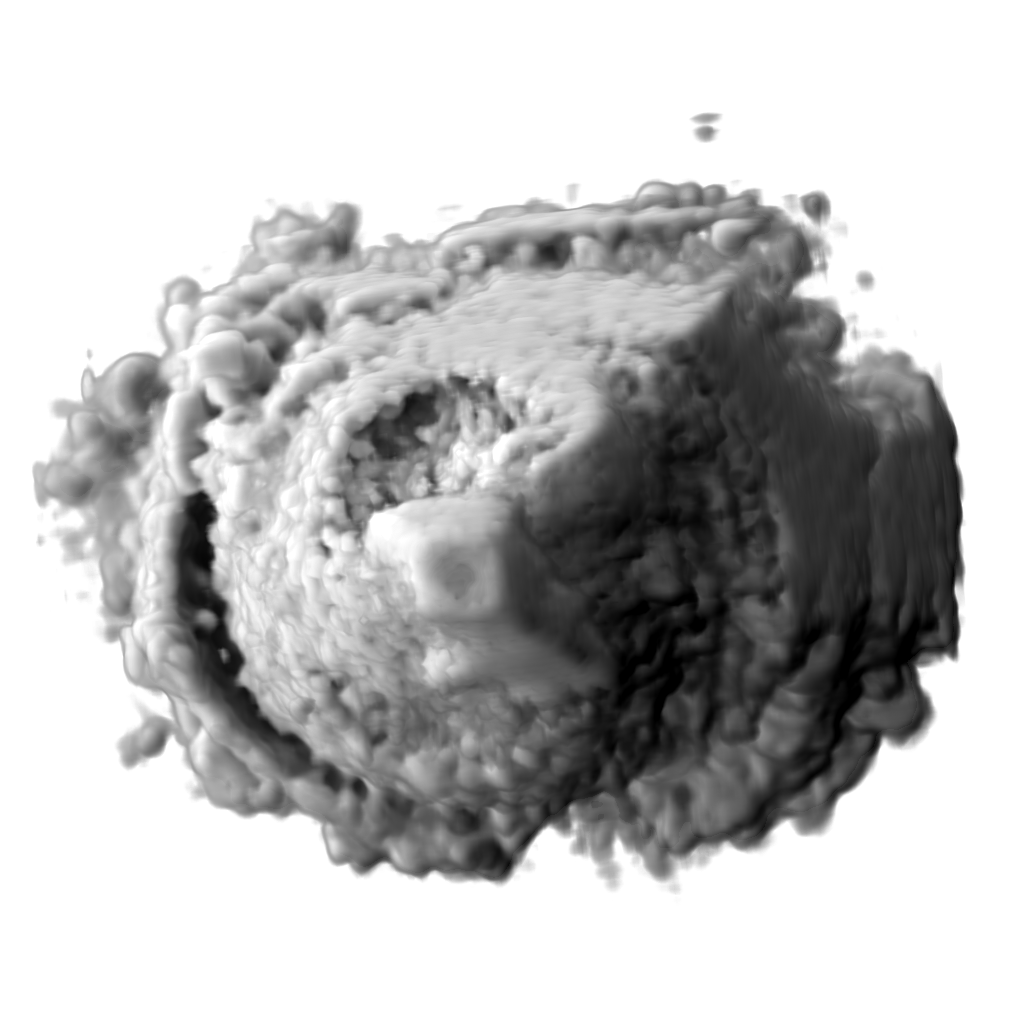}}
  \caption{Comparison of six volume shading models as applied to a 3D US scan of a human heart.}
  \label{fig:comparison}
\end{figure}

Recent works show that global illumination models based on gradient-free methods are suitable for rendering ultrasound volumes~\cite{Ropinski10,solteszova10}. Ropinski et al. described a volumetric lighting model which simulates scattering and shadowing~\cite{Ropinski10}. They use slice-based volume rendering from the view of the light source to calculate a light volume and raycasting to render the final image (see Figure~\ref{fig:hearttimo}). A perceptual evaluation of the generated images indicates, that the proposed model yields stronger depth cues than gradient-based shading. \v{S}olt\'{e}szov\'{a} et al. presented a single-pass method for simulation of light scattering in volumes~\cite{solteszova10}. Light transport is approximated using a tilted cone-shaped function which leaves elliptic footprints in the opacity buffer during slice-based volume rendering. They use a slice-based renderer with an additional opacity buffer. This buffer is incrementally blurred with an elliptical kernel, and the algorithm generates a high-quality soft-shadowing effect (see Figure~\ref{fig:heartme}). The light position and direction can be interactively modified. While these two techniques have been explicitly applied to 3D US data, the application of other volumetric illumination models potentially also improves the visual interpretation of 3D US data. Figure~\ref{fig:comparison} shows a comparison of six different shading techniques as applied to a 3D US scan of a human heart. While the first row of Figure~\ref{fig:comparison} shows examples for the already addressed shading techniques, the second row shows three alternative approaches. Figure~\ref{fig:comparison:half} incorporates scattering of light in volume data, as proposed by Kniss et al.~\cite{kniss02halfangle}. Their slicing technique allows textured slices to be rendered from both light and viewing direction simultaneously. By sampling the incident light from multiple directions while updating the light's attenuation map, they account for scattering effects in slice-based volume rendering. Figure~\ref{fig:comparison:dirocc} shows the application of the directional occlusion shading technique~\cite{schott09directionalocclusion}. This technique constrain the light source position to coincide with the view point. Finally, Figure~\ref{fig:comparison:sh} shows the application of a technique based on spherical harmonic lighting~\cite{lindemann10materials}.

Advanced illumination techniques are being now implemented in the commercial ultrasound workstations. Some workstations use additional color coding based on depth. Deeper tissues are colored with cold tones such as blue while close regions have red and orange tones. This effect has been firstly described by Einthoven \cite{Einthoven1885} and is also referred to as chromostereopsis \cite{Allen81}. Figure~\ref{fig:chromadepth} shows a chromatic depth-encoding rendering of a 3D human heart in a modern ultrasound workstation.

\section{Ultrasound and Augmented Reality}

Ultrasound is commonly viewed on a separate monitor. Therefore it is difficult to comprehend the spatial relationship between what you see on the monitor and where it is located in the patient's body. Augmented reality can aid the user by for instance super-imposing the ultrasound image onto the body where the ultrasound probe is positioned. 
Bajura et al. presented a system which linked 3D freehand ultrasound with a head-mounted display (HMD) \cite{bajura92}. The HMD contains a camera, tracker and two displays, one for each eye. The system can then project the tracked ultrasound image onto the tracked camera feed so the user can see where in the body the image actually is positioned.

Combining segmentation, surface rendering and augmented reality, Sato et al. aimed to aid surgeons during breast tumor removal for minimizing risk and maximizing breast preservation\cite{sato98} by projecting a segmented tumor onto a video feed. The tumor is segmented using a minimal intensity projection based selection of the volume of interest. In the final stage, the tumor is surface rendered and superimposed on the video image. 

Stetten et al. show how tomographic reflection can provide a superimposed image onto the body without any tracking systems\cite{stetten00}. The ultrasound probe carries a half-silvered mirror. The mirror reflects the ultrasound image which is shown on a flat panel monitor mounted on the probe. 
This technique was extended in the Sonic Flashlight \cite{shelton02}. The tomographic reflection showed to increase the localization perception compared to conventional ultrasound \cite{wu05}.


Augmented reality show great potential benefit in medical ultrasound imaging. Yet, there is a lag from technology development to the actual integration into the every day usage. Sielhorst et al. published a detailed review for advanced medical displays in 2008~\cite{sielhorst08}. This paper discuss the potential benefit and the increasing use for augmented reality in medical imaging as a whole. Stating that improvements in both technology is needed to be able to create a seamless integration into the physicians and surgeons work flow. 

\section{Summary and Discussion}
Medical ultrasound data is very different compared to other medical imaging modalities. Techniques for the different steps in the visualization pipeline are especially tailored to suit the different nature of the data acquisition. Techniques meant for in-vivo use have strong performance requirements to handle the high frame rate of ultrasound images. Yet, there is a great desire for techniques to improve, e.g., communication and training from our local medical partners. Research in advanced techniques focus greatly on 3D ultrasound, but the trend in diagnostics is mostly 2D due to higher frame-rate, high resolution and a minimal requirement for interaction. The temporal and spatial resolution for ultrasound is approaching the physical limit. The responsibility now lies on visualization techniques to take it further, combining high resolution 2D and contextual 3D ultrasound.

%
%





\bibliographystyle{plain}
\bibliography{star}

%

%




\end{document}